\documentclass{PoS}

\usepackage{amsmath}
\usepackage{latexsym}
\usepackage{amssymb}
\usepackage{dsfont}

\newcommand{\ud}{\mathrm{d}}

\newcommand{\uvec}[1]{\vec{#1}}
\newcommand{\pure}{\text{pure}}
\newcommand{\phys}{\text{phys}}

\title{Orbital angular momentum in the nucleon}

\ShortTitle{OAM in the nucleon}

\author{\speaker{C\'edric Lorc\'e}\\
        IFPA,  AGO Department, Universit\'e de Li\` ege, Sart-Tilman, 4000 Li\`ege, Belgium\\
        E-mail: \email{C.Lorce@ulg.ac.be}}

\abstract{In the last decade, it has been realized that the orbital angular momentum of partons inside the nucleon plays a major role. It contributes significantly to nucleon properties and is at the origin of many asymmetries observed in spin physics. It is therefore of paramount importance to determine this quantity if we want to understand the nucleon internal structure and experimental observables. This triggered numerous discussions and controversies about the proper definition of orbital angular momentum and its extraction from experimental data. We summarize the present situation and discuss recent developments in this field.}

\FullConference{XXII. International Workshop on Deep-Inelastic Scattering and Related Subjects,\\
		28 April - 2 May 2014\\
		Warsaw, Poland}

\begin{document}

\section{Introduction}	

Unlike familiar quantum bound states, nucleons are strongly affected by relativistic effects, making their description particularly complicated. Nevertheless, it is possible to decompose their total angular momentum into spin and orbital angular momentum (OAM) contributions of their constituents, \emph{i.e.} quarks and gluons.

At the classical level, there already exist two definitions of momentum: canonical and kinetic. In many systems, both definitions coincide. But in the presence of gauge fields, these two definitions differ. It is therefore not surprising that there correspond two definitions of OAM, and therefore two decompositions of total angular momentum. Deciding which quantities, canonical or kinetic, have to be considered as the ``physical'' ones has a long history and remains the source of intense debates. On the one hand, the canonical decomposition is complete and has a clear interpretation in terms of quark and gluon contributions, but is apparently not gauge invariant. On the other hand, the kinetic decomposition is manifestly gauge invariant, but is incomplete (in the sense that there is no notion of spin and OAM for a gauge particle) and has a less clear interpretation in terms of quark and gluon contributions.

These problems can basically be cured by the notion of gauge-invariant extension (GIE). The idea is to find a manifestly gauge-invariant quantity which coincides with a gauge non-invariant quantity in a particular gauge. This allows one to extend the validity of a physical interpretation considered in a certain gauge to any gauge. The price to pay for such an approach is that GIEs are usually non-local expressions of the gauge fields. This non-locality does not violate causality since it can be removed by a simple gauge transformation. The only problem is that to any gauge non-invariant quantity, one can in theory associate an \emph{infinite} number of GIEs, just by changing the gauge where they are required to coincide. This last issue is solved by examining how one experimentally probes the internal structure of the nucleon. It is indeed the physical process and its theoretical description which dictate at the end what is the most convenient or natural GIE to use.

The crucial question now is how to connect these definitions of OAM to measurable quantities. Over the last two decades or so, many relations and sum rules have been proposed. Some are exact, while others are suggested by rather simple models of the nucleon structure. On the experimental side, a lot of effort is currently invested in the measurement of these quantities. On the theoretical side, people are working hard to understand the difference between the various definitions and to identify physical observables sensitive to OAM at a quantitative level.

In this proceeding, we sketch a summary of the present situation and discuss some recent developments. In section \ref{sec2}, we present the two families of proton spin decompositions. In section \ref{sec3}, we discuss the GIE approach and the related uniqueness issue. In section \ref{sec4}, we summarize how OAM and quark spin-orbit correlation can be expressed in terms of parton distributions. Finally, we conclude with section \ref{sec5}. For the interested reader, more detailed discussions of the topic can be found in the recent review~\cite{Leader:2013jra}.

\section{Canonical and kinetic spin decompositions}\label{sec2}

The so-called canonical decomposition reads
\begin{equation}
\uvec J=\uvec S^q+\uvec \ell^q+\uvec S^G+\uvec \ell^G.
\end{equation}
The explicit expressions for the quark/gluon spin/OAM contributions were given by Jaffe and Manohar~\cite{Jaffe:1989jz}
\begin{equation}
\begin{aligned}
\uvec S^q&=\int\ud^3x\,\psi^\dag\tfrac{1}{2}\uvec\Sigma\psi, \qquad&\uvec \ell^q_\text{JM}&=\int\ud^3x\,\psi^\dag(\uvec x\times\tfrac{1}{i}\uvec\nabla)\psi,\\
\uvec S^G_\text{JM}&=\int\ud^3x\,\uvec E^a\times\uvec A^a,&\uvec \ell^G_\text{JM}&=\int\ud^3x\,E^{ai}(\uvec x\times\uvec\nabla) A^{ai}.
\end{aligned}
\end{equation}
Except for the quark spin contribution $\uvec S^q$, the other contributions are not gauge invariant. In order to make sense of it, Jaffe and Manohar stressed that this decomposition has to be considered in a fixed gauge. Later, Chen \emph{et al.}~\cite{Chen:2008ag,Chen:2009mr} managed to write down the corresponding GIE
\begin{equation}\label{gic}
\begin{aligned}
\uvec S^q&=\int\ud^3x\,\psi^\dag\tfrac{1}{2}\uvec\Sigma\psi, \qquad&\uvec \ell^q_\text{GIE}&=\int\ud^3x\,\psi^\dag(\uvec x\times i\uvec D_\pure)\psi,\\
\uvec S^G_\text{GIE}&=\int\ud^3x\,\uvec E^a\times\uvec A^a_\phys,&\uvec \ell^G_\text{GIE}&=-\int\ud^3x\,E^{ai}(\uvec x\times\uvec{\mathcal D}^{ab}_\pure) A^{bi}_\phys,
\end{aligned}
\end{equation}
by decomposing the gauge field into two parts $\vec A=\vec A^\pure+\vec A^\phys$ and using only $\vec A^\pure$ in the covariant derivatives $\uvec D_\pure=-\uvec\nabla-ig\uvec A_\pure$ and $\uvec{\mathcal D}_\pure=-\uvec\nabla-ig[\uvec A_\pure,\quad]$.

The so-called kinetic decomposition reads
\begin{equation}
\uvec J=\uvec S^q+\uvec L^q+\uvec S^G+\uvec L^G.
\end{equation}
The corresponding explicit expressions for the quark and gluon contributions were given by Ji~\cite{Ji:1996ek}
\begin{equation}
\begin{aligned}
\uvec S^q&=\int\ud^3x\,\psi^\dag\tfrac{1}{2}\uvec\Sigma\psi, \qquad&\uvec L^q&=\int\ud^3x\,\psi^\dag(\uvec x\times i\uvec D)\psi,\\
\uvec J^G&=\uvec S^G+\uvec L^G=\int\ud^3x\,\uvec x\times(\uvec E^a\times\uvec B^a).\quad
\end{aligned}
\end{equation}
In accordance with textbook claims, Ji was not able to write down a gauge-invariant expression separately for gluon spin and OAM contributions. Later, Wakamatsu~\cite{Wakamatsu:2010qj,Wakamatsu:2010cb} used the Chen \emph{et al.} approach to make the kinetic decomposition complete
\begin{equation}\label{gik}
\begin{aligned}
\uvec S^q&=\int\ud^3x\,\psi^\dag\tfrac{1}{2}\uvec\Sigma\psi, \qquad&\uvec L^q&=\int\ud^3x\,\psi^\dag(\uvec x\times i\uvec D)\psi,\\
\uvec S^G_\text{GIE}&=\int\ud^3x\,\uvec E^a\times\uvec A^a_\phys,&\uvec L^G_\text{GIE}&=\uvec J^G-\uvec S^G_\text{GIE},
\end{aligned}
\end{equation}

At the end of the day, the sole difference between the complete gauge-invariant canonical~\eqref{gic} and kinetic~\eqref{gik} decompositions is in the interpretation of the so-called potential OAM as either quark or gluon contribution
\begin{equation}
\begin{aligned}
\uvec L_\text{pot}&=\uvec L^q-\uvec \ell^q_\text{GIE}=\int\ud^3x\,\psi^\dag (\uvec x\times g\uvec A_\phys)\psi,&\\
&=-(\uvec L^G_\text{GIE}-\uvec \ell^G_\text{GIE})=\int\ud^3x\,(\uvec{\mathcal D}\cdot\uvec E)^a\,\uvec x\times\uvec A^a_\phys,
\end{aligned}
\end{equation}
using the QCD equation of motion $(\uvec{\mathcal D}\cdot\uvec E)^a=g\psi^\dag t^a\psi$.

\section{Chen \emph{et al.} approach and GIE}\label{sec3}

The complete gauge-invariant decompositions~\eqref{gic} and~\eqref{gik} seem to contradict textbook claims that it is not possible to write down gauge-invariant expressions for gluon spin and OAM contributions. There is no genuine contradiction since textbooks actually refer to local expressions only. The Chen \emph{et al.} approach circumvent the impossibility by considering intrinsically non-local expressions.

The Chen \emph{et al.} approach is an explicit realization of the GIE concept. In the covariant form, it consists in a splitting the gauge potential into ``pure-gauge'' and ``physical'' terms~\cite{Chen:2008ag,Chen:2009mr,Wakamatsu:2010qj,Wakamatsu:2010cb,Lorce:2012rr}
\begin{equation}\label{decomposition}
A_\mu=A^\pure_\mu+A^\phys_\mu,
\end{equation}
where, by definition, the pure-gauge term does not contribute to the field strength
\begin{equation}\label{cond1}
F^\pure_{\mu\nu}=\partial_\mu A^\pure_\nu-\partial_\nu A^\pure_\mu-ig[A^\pure_\mu,A^\pure_\nu]=0
\end{equation}
and transforms like $A_\mu$
\begin{equation}\label{cond2}
A^\pure_\mu\mapsto \tilde A^\pure_\mu=U[A^\pure_\mu+\tfrac{i}{g}\partial_\mu]U^{-1}
\end{equation}
under gauge transformations. Consequently, the physical term is responsible for the field strength
\begin{equation}
F_{\mu\nu}=\mathcal D^\pure_\mu A^\phys_\nu-\mathcal D^\pure_\nu A^\phys_\mu-ig[A^\phys_\mu,A^\phys_\nu],
\end{equation}
and transforms like the latter
\begin{equation}
A^\phys_\mu\mapsto \tilde A^\phys_\mu=UA^\phys_\mu U^{-1}
\end{equation}
under gauge transformations. This approach turns out to be very similar to the background field method~\cite{Lorce:2013bja} and is closely related to the so-called Dirac variables~\cite{Chen:2012vg,Lorce:2013gxa}.

The main problem with the Chen \emph{et al.} approach is that the splitting into pure-gauge and physical terms is not unique. Indeed, the following alternative fields
\begin{equation}\label{Stueckelberg}
\bar A^\pure_\mu=A^\pure_\mu+B_\mu,\qquad\bar A^\phys_\mu=A^\phys_\mu-B_\mu,
\end{equation}
also satisfy the defining conditions~\eqref{cond1} and~\eqref{cond2}, provided that $B_\mu$ transforms in a suitable way under gauge transformations~\cite{Lorce:2012rr,Stoilov:2010pv}. Since the pure-gauge term plays essentially the role of a background field, the freedom~\eqref{Stueckelberg} simply corresponds to background dependence~\cite{Lorce:2013bja}. It can also be understood from a non-local point of view, where $A_\mu^\pure$ and $A_\mu^\phys$ appear as particular functionals of $A_\mu$~\cite{Hatta:2011zs,Hatta:2011ku,Lorce:2012ce}. But as already mentioned in the introduction, it is basically the actual physical process and its theoretical description which determine the most convenient or natural GIE to use, \emph{i.e.} which constraint to impose on $A^\phys_\mu$.

\section{OAM and spin-orbit correlation}\label{sec4}

The most natural expression for OAM is as a phase-space integral \cite{Lorce:2011kd,Lorce:2011ni}
\begin{equation}\label{OAMWigner}
l_z^q(\mathcal W)=\int\ud x\,\ud^2k_\perp\,\ud^2b_\perp\,(\vec b_\perp\times\vec k_\perp)_z\,\rho^{[\gamma^+]q}_{++}(x,\vec k_\perp,\vec b_\perp;\mathcal W),
\end{equation}
where the relativistic phase-space or Wigner distribution $\rho^{[\gamma^+]q}_{++}(x,\vec k_\perp,\vec b_\perp;\mathcal W)$ can be interpreted as giving the quasi-probability to find an unpolarized quark with longitudinal momentum fraction $x$, transverse momentum $\uvec k_\perp$ and transverse position $\uvec b_\perp$ inside a longitudinally polarized nucleon. The phase-space distributions can be expressed in terms of so-called generalized transverse-momentum dependent distributions (GTMDs)~\cite{Meissner:2009ww,Lorce:2013pza}, leading to the simple relation~\cite{Hatta:2011ku,Lorce:2011kd,Kanazawa:2014nha}
\begin{equation}
l_z^q(\mathcal W)=-\int\ud x\,\ud^2k_\perp\,\tfrac{\uvec k^2_\perp}{M^2}\,F^q_{14}(x,0,\uvec k_\perp,\uvec 0_\perp;\mathcal W).
\end{equation}
This OAM depends on the type of Wilson line $\mathcal W$ involved in the definition of the non-local quark correlator~\cite{Lorce:2012ce,Ji:2012sj,Burkardt:2012sd}. The (light-front) canonical OAM is obtained with a (light-front) staple-like Wilson line $\ell^q_z=l^q_z(\mathcal W_\text{staple})$, while the kinetic OAM is obtained with a direct straight Wilson line $L^q_z=l^q_z(\mathcal W_\text{straight})$. Unfortunately, it is not known so far how to extract quark GTMDs from actual experiments. There exist however alternative relations in terms of measurable distributions. The kinetic OAM can be expressed in terms of twist-2~\cite{Ji:1996ek} and twist-3~\cite{Penttinen:2000dg} generalized parton distributions (GPDs)
\begin{align}
L^q_z&=\frac{1}{2}\int\ud x\left\{x\left[H^q(x,0,0)+E^q(x,0,0)\right]-\tilde H^q(x,0,0)\right\}\\
&=-\int\ud x\,xG^q_2(x,0,0).
\end{align}
For the canonical OAM, the following relation to transverse-momentum dependent distributions (TMDs) has been proposed
\begin{equation}
\mathcal L_z^q=-\int\ud x\,\ud^2k_\perp\,\tfrac{\uvec k_\perp^2}{2M^2}\,h_{1T}^{\perp q}(x,\uvec k^2_\perp),
\end{equation}
but is valid in some quark models only~\cite{Lorce:2011kn}. 

Adding an extra $\gamma_5$ matrix to the quark OAM operator gives the quark spin-orbit operator $c^q_z$ which measures the correlation between the quark spin and the quark OAM~\cite{Lorce:2011kd,Lorce:2014mxa}. Like the OAM, its most natural expression is as a phase-space integral
\begin{align}
c_z^q(\mathcal W)&=\int\ud x\,\ud^2k_\perp\,\ud^2b_\perp\,(\vec b_\perp\times\vec k_\perp)_z\,\rho^{[\gamma^+\gamma_5]q}_{++}(x,\vec k_\perp,\vec b_\perp;\mathcal W),\\
&=\int\ud x\,\ud^2k_\perp\,\tfrac{\uvec k^2_\perp}{M^2}\,G^q_{11}(x,0,\uvec k_\perp,\uvec 0_\perp;\mathcal W),
\end{align}
and depends on the type of Wilson line $\mathcal W$ involved in the definition of the non-local quark correlator.  The quark kinetic spin-orbit correlation $C^q_z=c^q_z(\mathcal W_\text{straight})$ can also be expressed in terms of twist-2 and twist-3 GPDs~\cite{Lorce:2014mxa}
\begin{align}
C^q_z&=\frac{1}{2}\int\ud x\left[x\tilde H^q(x,0,0)- H^q(x,0,0)\right]+\mathcal O(\tfrac{m_q}{M})\\
&=-\int\ud x\,x\left[\tilde G^q_2(x,0,0)+2\tilde G^q_4(x,0,0)\right].
\end{align}

\section{Conclusion}\label{sec5}

There are essentially two (families of) proton spin decompositions: canonical and kinetic. The concept of gauge-invariant extension ensures that the different contributions are in principle measurable, in the sense that they can be defined in a manifestly gauge-invariant way. The price to pay is that the corresponding operators are usually non-local, leading to a uniqueness problem, which is however solved by the framework used to describe the actual experiment. Quark orbital angular momentum and spin-orbit correlations are most naturally defined in terms of phase-space distributions, but are experimentally accessible \emph{via} other relations to measurable parton distributions.

\section*{Acknowledgements}
I benefited a lot from many discussions and collaborations with E. Leader, B. Pasquini and  M. Wakamatsu. This work was supported by the Belgian Fund F.R.S.-FNRS \emph{via} the contract of Charg\'e de Recherches.



\begin{thebibliography}{00}  

\bibitem{Leader:2013jra} 
  E.~Leader and C.~Lorc\'e,
  to appear in Phys.\ Rept. (2014),
  arXiv:1309.4235 [hep-ph].

\bibitem{Jaffe:1989jz} 
  R.~L.~Jaffe and A.~Manohar,
  Nucl.\ Phys.\ B {\bf 337}, 509 (1990).

\bibitem{Chen:2008ag} 
  X.~-S.~Chen, X.~-F.~Lu, W.~-M.~Sun, F.~Wang and T.~Goldman,
  Phys.\ Rev.\ Lett.\  {\bf 100}, 232002 (2008).

\bibitem{Chen:2009mr} 
  X.~-S.~Chen, W.~-M.~Sun, X.~-F.~Lu, F.~Wang and T.~Goldman,
  Phys.\ Rev.\ Lett.\  {\bf 103}, 062001 (2009).

\bibitem{Ji:1996ek} 
  X.~-D.~Ji,
  Phys.\ Rev.\ Lett.\  {\bf 78}, 610 (1997).

\bibitem{Wakamatsu:2010qj} 
  M.~Wakamatsu,
  Phys.\ Rev.\ D {\bf 81}, 114010 (2010).

\bibitem{Wakamatsu:2010cb} 
  M.~Wakamatsu,
  Phys.\ Rev.\ D {\bf 83}, 014012 (2011).

\bibitem{Lorce:2012rr} 
  C.~Lorc\'e,
  Phys.\ Rev.\ D {\bf 87}, 034031 (2013).

\bibitem{Lorce:2013bja} 
  C.~Lorc\'e,
  Nucl.\ Phys.\ A {\bf 925}, 1 (2014).

\bibitem{Chen:2012vg} 
  X.~-S.~Chen,
  arXiv:1203.1288 [hep-ph].

\bibitem{Lorce:2013gxa} 
  C.~Lorc\'e,
  Phys.\ Rev.\ D {\bf 88}, 044037 (2013).

\bibitem{Stoilov:2010pv} 
  M.~N.~Stoilov,
  arXiv:1011.5617 [hep-th].

\bibitem{Hatta:2011zs} 
  Y.~Hatta,
  Phys.\ Rev.\ D {\bf 84}, 041701 (2011).

\bibitem{Hatta:2011ku} 
  Y.~Hatta,
  Phys.\ Lett.\ B {\bf 708}, 186 (2012).

\bibitem{Lorce:2012ce} 
  C.~Lorc\'e,
  Phys.\ Lett.\ B {\bf 719}, 185 (2013).

\bibitem{Lorce:2011kd} 
  C.~Lorc\'e and B.~Pasquini,
  Phys.\ Rev.\ D {\bf 84}, 014015 (2011).

\bibitem{Lorce:2011ni} 
  C.~Lorc\'e, B.~Pasquini, X.~Xiong and F.~Yuan,
  Phys.\ Rev.\ D {\bf 85}, 114006 (2012).

\bibitem{Meissner:2009ww} 
  S.~Meissner, A.~Metz and M.~Schlegel,
  JHEP {\bf 0908}, 056 (2009).

\bibitem{Lorce:2013pza} 
  C.~Lorc\'e and B.~Pasquini,
  JHEP {\bf 1309}, 138 (2013).

\bibitem{Kanazawa:2014nha} 
  K.~Kanazawa, C.~Lorc\'e, A.~Metz, B.~Pasquini and M.~Schlegel,
  to appear in PRD (2014),
  arXiv:1403.5226 [hep-ph].

\bibitem{Ji:2012sj} 
  X.~Ji, X.~Xiong and F.~Yuan,
  Phys.\ Rev.\ Lett.\  {\bf 109}, 152005 (2012).

\bibitem{Burkardt:2012sd} 
  M.~Burkardt,
  Phys.\ Rev.\ D {\bf 88}, 014014 (2013).

\bibitem{Penttinen:2000dg} 
  M.~Penttinen, M.~V.~Polyakov, A.~G.~Shuvaev and M.~Strikman,
  Phys.\ Lett.\ B {\bf 491}, 96 (2000).

\bibitem{Lorce:2011kn} 
  C.~Lorc\'e and B.~Pasquini,
  Phys.\ Lett.\ B {\bf 710}, 486 (2012).

\bibitem{Lorce:2014mxa} 
  C.~Lorc\'e,
 to appear in PLB (2014),
  arXiv:1401.7784 [hep-ph].



\end{thebibliography}
\end{document}